\documentclass{article}
\usepackage{spconf,amsmath,graphicx,hyperref}
\usepackage{booktabs}
\usepackage{multirow} 
\usepackage{amssymb}
\usepackage{cite}
\usepackage{amsmath,amsfonts}
\usepackage{algorithmic}
\usepackage{algorithm}
\usepackage{array}
\usepackage[caption=false,font=normalsize,labelfont=sf,textfont=sf]{subfig}
\usepackage{tabularray}
\usepackage{textcomp}
\usepackage{stfloats}
\usepackage{url}
\usepackage{verbatim}
\usepackage{graphicx}
\usepackage{cite}
\usepackage{enumitem}
\usepackage{hyperref}
\usepackage{makecell}
\usepackage{etoolbox}
\robustify\bfseries
\usepackage{amssymb}
\usepackage{tabularx}
\usepackage{wasysym}
\usepackage{color,soul}
\usepackage{balance}
\usepackage{booktabs}
\usepackage{multirow}
\usepackage{siunitx}
\usepackage{arydshln} 
\usepackage{tabularray}
\usepackage{tblr-extras}
\def\H{{\mathsf H}}

\usepackage{siunitx}
\usepackage{booktabs}
\usepackage{caption}
\usepackage{flushend}

\title{VM-ArrayDPS: Virtual Microphone Augmented Diffusion Posterior Sampling for Unsupervised Blind Speech Separation}

\name{Jingqi Sun, Haozhan Tang, Shulin He, and Zhong-Qiu Wang}
\address{Department of Computer Science and Engineering\\
Southern University of Science and Technology, Shenzhen, China\\
\texttt{\small jingqi.sun@outlook.com, wang.zhongqiu41@gmail.com}
}

\makeatletter
\renewcommand\normalsize{%
   \@setfontsize\normalsize{9pt}{10.5pt}} 
\makeatother
\let\OLDthebibliography\thebibliography
\renewcommand\thebibliography[1]{
  \OLDthebibliography{#1}
  \setlength{\parskip}{0.5pt}
  \setlength{\itemsep}{1pt plus 0.3ex}
}
\makeatletter
\providecommand{\bstctlcite}[1]{%
  \@bsphack
  \if@filesw\immediate\write\@auxout{\string\citation{#1}}\fi
  \@esphack}
\makeatother

\begin{document}
\bstctlcite{IEEEAuthorControl}
\maketitle

\begin{abstract}
Blind source separation (BSS) aims to separate multiple source signals from their mixtures without prior knowledge of the sources or mixing process.
ArrayDPS addresses this task using a pretrained speech diffusion model, with multi-channel consistency (MC) constraints derived from physical microphone observations guiding posterior sampling.
However, these constraints become less effective when only a few physical microphones are available and their recordings have low signal-to-noise ratios (SNRs).
To address these limitations,  we propose VM-ArrayDPS, which uses independent vector analysis (IVA)-based spatial demixing to construct higher-SNR virtual microphone observations from the physical recordings.
These virtual observations are incorporated alongside the physical microphone signals through a weighted MC objective, allowing the IVA outputs to guide source recovery throughout posterior sampling.
Evaluation results on 2- and 3-speaker separation in reverberant conditions demonstrate consistent improvements over ArrayDPS.
\end{abstract}


\section{Introduction}
\label{sec:intro}

Separating overlapping speech recorded by microphone arrays is challenging when only a few microphones are available and the recordings contain noise and reverberation.
When neither the source signals nor the mixing process is known, this task is referred to as blind speech separation (BSS) \cite{cao1996general,sawada2019review}.
Classical approaches such as independent vector analysis (IVA) \cite{kim2007blind,kim2006independent,lee2007fast,ono2011stable,kitamura2016determined,luo2023independent} address this task by exploiting statistical independence across sources and dependencies across frequency bins.
Although these filters enhance individual sources, the outputs may still contain residual interference and noise.

Beyond the statistical source models used in classical methods, neural speech separation methods learn speech structure directly from data \cite{wang2018supervised}.
Supervised methods typically learn a direct mapping from mixtures to clean sources, requiring paired mixture--source training data, and their performance may degrade when the test conditions differ from those encountered during training \cite{luo2019conv,luo2020dual,wang2023tf,hershey2016deep,kolbaek2017multitalker,vzmolikova2019speakerbeam,wang2019deep,liu2019divide,wang2018combining,luo2018speaker,yu2017permutation,chen2017deep}.
Multi-channel speech separation further exploits spatial information from microphone arrays to distinguish and recover individual sources \cite{luo2019fasnet,wang2018multi,tesch2023multi,luo2020end,chen2021beam}. 
Unsupervised approaches such as UNSSOR remove the need for clean source references by exploiting mixture consistency during training, but still learn a task-specific separation model from the training mixture distribution \cite{wang2023unssor,wisdom2020unsupervised}.
An alternative is to use a pretrained speech diffusion model to infer the sources at test time, without training a task-specific separator\cite{scheibler2023diffusion,lutati2023separate}.
To separate a particular mixture, the sampling process must be guided by the recorded signals.
Diffusion posterior sampling (DPS) provides such guidance through the gradient of a measurement log-likelihood, which evaluates how well the source estimates account for the observations under a given measurement model\cite{chung2022diffusion}.
In blind multi-channel separation, the acoustic filters defining this model are unknown.
ArrayDPS estimates these filters from the current source estimates and microphone recordings, and constructs a multi-channel consistency (MC) objective from the resulting reconstruction error to guide posterior sampling\cite{xu2025arraydps}.

This observation-based guidance, however, is constrained by the quality and number of available microphone observations. ArrayDPS uses forward convolutive prediction (FCP) \cite{wang2021convolutive,wang2021convolutiveconf} to estimate acoustic filters for reconstruction. With few microphones, the resulting MC objective provides limited spatial constraints on source recovery. Low-SNR recordings can further degrade acoustic filter estimation, making the reconstruction error less reliable for guiding posterior sampling. Together, these limitations weaken the observation-based guidance needed for accurate separation.

To address these limitations, we propose VM-ArrayDPS, which augments physical microphone observations with source-enhanced virtual microphones to strengthen the guidance for posterior sampling.
These virtual observations are generated through IVA-based spatial demixing, which suppresses interfering sources and produces signals with higher SNR \cite{kim2006independent}.
While ArrayDPS uses IVA outputs only for warm initialization \cite{xu2025arraydps}, VM-ArrayDPS incorporates them as virtual microphone observations throughout posterior sampling.
FCP estimates convolutional mappings from the source estimates to these virtual observations, and the corresponding reconstruction errors contribute additional terms to the MC objective.
By introducing source-enhanced observations, VM-ArrayDPS supplements the limited physical observations and reduces reliance on their noisy reconstruction constraints.

Based on this observation, we propose \textbf{VM-ArrayDPS}, which augments ArrayDPS with higher-SNR virtual microphone observations without additional training or clean separation targets. The main contributions are:
\begin{itemize}[leftmargin=*,noitemsep,topsep=0pt]
    \item We introduce virtual microphone augmentation into ArrayDPS by incorporating IVA-derived observations as additional measurements for posterior sampling.
    \item We formulate a virtual-microphone MC objective and introduce a weighting strategy to control the contribution of physical and virtual observations.
\end{itemize}
Experiments on 2- and 3-speaker SMS-WSJ \cite{drude2019sms} demonstrate consistent improvements over ArrayDPS \cite{xu2025arraydps}, with ablations analyzing the effects of the number and weight of virtual microphones.

\section{Review of ArrayDPS}
\label{sec:background}


Consider a mixture of $C$ reverberant speech sources observed by $M$ microphones:
\begin{equation}
y_m(t)=\sum_{c=1}^{C} h_{m,c}(t)*s_c(t)+n_m(t),
\label{eq:mix}
\end{equation}
where $s_c(t)$ is the $c$-th source signal, $h_{m,c}(t)$ is its room impulse response (RIR) to microphone $m$, $*$ denotes convolution, and $n_m(t)$ denotes additive noise. In blind separation, neither the sources nor the acoustic filters are assumed known.

ArrayDPS \cite{xu2025arraydps} formulates separation as sampling from the posterior distribution
\begin{equation}
p(s_{1:C}\mid y_{1:M}),
\label{eq:posterior}
\end{equation}
where $s_{1:C}$ and $y_{1:M}$ denote all source and microphone signals, respectively. 
DPS \cite{chung2022diffusion} combines a learned prior with a measurement likelihood. In the score-based formulation used by ArrayDPS, the posterior score can be decomposed by Bayes' rule as
\begin{equation}
\nabla_{\mathbf{s}^{\tau}}\log p(s^{\tau}\mid y)
=
\nabla_{s^{\tau}}\log p(s^{\tau})
+
\nabla_{s^{\tau}}\log p(y\mid s^{\tau}),
\label{eq:score}
\end{equation}
where $\tau$ is the diffusion noise level.
The first term is the prior score. A pretrained speech diffusion model supplies a denoised estimate $\hat{s}^{\tau}$, from which the score can be estimated using Tweedie's formula \cite{efron2011tweedie}:
\begin{equation}
\nabla_{s^{\tau}}\log p(s^{\tau})
=
\frac{\hat{s}^{\tau}-s^{\tau}}{\sigma(\tau)^2}.
\label{eq:tweedie}
\end{equation}
This prior encourages the recovered signals to remain on the manifold of plausible speech.
To connect the source prior to the actual observations, ArrayDPS constructs an estimate of each microphone signal using estimated acoustic transfer functions:
\begin{equation}
\hat{y}_m(t)=\sum_{c=1}^{C}(\hat{h}_{m,c}*\hat{s}_c^\tau)(t).
\label{eq:recon}
\end{equation}
Assuming additive Gaussian observation noise, the negative log-likelihood is proportional to a reconstruction error,
\begin{equation}
-\log p(y\mid s^{\tau})
\propto
\sum_{m=1}^{M}
\left\|y_m-\hat{y}_m(s^{\tau})\right\|_2^2.
\label{eq:mc}
\end{equation}
Consequently, the likelihood score is implemented as the gradient of the MC reconstruction loss:
\begin{equation}
\nabla_{s^{\tau}}\log p(y\mid s^{\tau})
\approx
-\xi(\tau)
\nabla_{s^{\tau}}
\sum_{m=1}^{M}
\left\|y_m-\hat{y}_m(s^{\tau})\right\|_2^2,
\label{eq:likelihood}
\end{equation}
where $\xi(\tau)$ controls the likelihood contribution at the current diffusion level.


ArrayDPS estimates the acoustic transfer functions using FCP \cite{wang2021convolutive,wang2021convolutiveconf}.
In the short-time Fourier transform (STFT) domain, let $Y_m(\ell,f)$ denote the observation at microphone $m$, and $\hat{S}_c(\ell,f)$ denote the direct-path estimate of source $c$ at time frame $\ell$ and frequency $f$.
A $K$-tap frequency-dependent FCP filter $\hat{h}_{m,c}(f)$ is estimated by
\begin{equation}
\hat{h}_{m,c}(f)=
\arg\min_{h_{m,c}(f)}
\sum_\ell
\frac{
|Y_m(\ell,f)-h_{m,c}(f)^{\H}\tilde{\hat{S}}_c(\ell,f)|^2
}{
\hat{\lambda}_c(\ell,f)
},
\label{eq:fcp}
\end{equation}
where $\tilde{\hat{S}}$ stacks the required delayed frames and $\hat{\lambda}$ is an observation-dependent weighting term. The estimated filter is then used as a frequency-dependent approximation of the RTF for reconstructing the microphone observations.

Finally, ArrayDPS performs the sampling process with the EDM sampler \cite{karras2022elucidating}. The IVA output is used as a warm initialization, and the prior and likelihood scores are repeatedly combined to update the source estimates. This design allows a single-speaker speech diffusion model to be applied to the blind multi-speaker separation problem without supervised separation training \cite{xu2025arraydps}.

\section{VM-ArrayDPS}
\label{sec:method}

\begin{figure}[t]
    \centering
    \includegraphics[width=\linewidth]{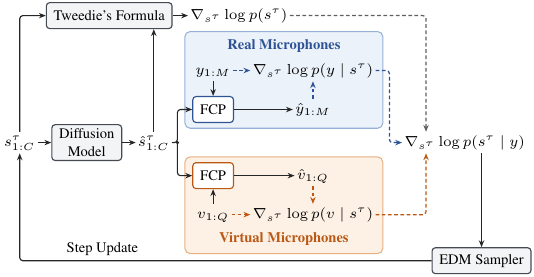}
    \caption{Gradient computation of VM-ArrayDPS at a diffusion step. Given the current source estimate $\mathbf{s}^{\tau}$, the physical and virtual microphone observations are reconstructed and used to compute the corresponding likelihood gradients.}
    \label{fig:vm_arraydps}
\end{figure}

VM-ArrayDPS strengthens DPS by incorporating source-enhanced virtual microphone observations.
Generated through IVA-based spatial demixing, these observations supplement the physical microphone signals and provide additional consistency constraints for source recovery.
Both physical and virtual observations are incorporated into a unified reconstruction objective, with a weighting factor controlling the contribution of the virtual observations.
Fig. \ref{fig:vm_arraydps} illustrates how this objective guides source estimation at each diffusion step.
The following subsections describe virtual microphone construction and its integration into posterior sampling.

\subsection{Virtual Microphone Construction}
Although the physical microphone observations provide spatial information, only a small number of microphones may be available in practice. 
This limits the number of observation constraints imposed on the source estimates, especially when multiple sources need to be recovered from the same set of mixtures. 
We therefore construct additional observations from the available microphone array, with the goal of obtaining constraints that are more source-oriented while remaining compatible with the forward acoustic model.

To this end, we apply IVA to the physical microphone signals and obtain separated signals through frequency-dependent spatial demixing filters \cite{kim2006independent}. Compared with the original microphone mixtures, these outputs are more source-oriented and typically exhibit a higher target-to-interference ratio. We use these separated signals as the basis for constructing virtual microphone observations.

We refer to the IVA outputs as \emph{virtual microphones} because each output can be viewed as a transformed observation of the same underlying acoustic sources. The term ``virtual'' emphasizes that these observations do not correspond to additional physical microphone positions. Instead, they are obtained by applying linear spatial transformations to the available physical microphone signals, and can therefore be incorporated into the same forward acoustic model. For an IVA output $v_j(t)$, we estimate a corresponding filter $\hat{g}_{j,c}$ and reconstruct it from the current source estimates:
\begin{equation}
\hat{v}_j(t)=
\sum_{c=1}^{C}(\hat{g}_{j,c}*s_c^\tau)(t).
\label{eq:virtual-recon}
\end{equation}
This construction allows the source-oriented information contained in the IVA outputs to impose additional constraints on the current source estimates without introducing additional physical microphones or clean training data.

\subsection{Virtual-Microphone Consistency}

The virtual microphones are incorporated into the observation likelihood by requiring the current source estimates to explain both the physical microphone signals and the virtual microphone signals. Let $v_{1:Q}$ denote $Q$ virtual microphone signals. Assuming conditional independence between physical and virtual observations given the source estimates, the likelihood score becomes
\begin{align}
\nabla_{s^{\tau}}
\log p(y,v\mid s^{\tau})
\approx{}&
-\xi(\tau)\nabla_{s^{\tau}}
\sum_{m=1}^{M}
\| y_m-\hat{ y}_m\|_2^2
\nonumber\\
&-\zeta(\tau)\nabla_{s^{\tau}}
\sum_{j=1}^{Q}
\|v_j-\hat{v}_j\|_2^2.
\label{eq:joint}
\end{align}

The physical and virtual observations provide complementary constraints: the former retain the original spatial mixtures, while the latter provide source-oriented projections derived from the same microphone array. However, the reliability of a virtual observation is not necessarily identical to that of a physical microphone. An IVA estimate can contain residual interference or artifacts, and forcing exact consistency with it may therefore be harmful. We therefore introduce a relative weight $\alpha$ and use
\begin{align}
\nabla_{s^{\tau}}
\log p(y,v\mid s^{\tau})
\approx{}&
-\xi(\tau)\nabla_{s^{\tau}}
\sum_{m=1}^{M}
\|y_m-\hat{y}_m\|_2^2
\nonumber\\
&-\alpha\cdot\xi(\tau)\nabla_{s^{\tau}}
\sum_{j=1}^{Q}
\|v_j-\hat{v}_j\|_2^2.
\label{eq:weighted}
\end{align}
The original ArrayDPS likelihood is recovered when $Q=0$. In our experiments, $\alpha$ is tuned on the validation setting. Given $M$ physical microphones and $C$ sources, the construction can generate up to $M\cdot C$ virtual observations. In practice, we use only a small number of virtual microphones to improve the observation constraints while limiting the additional computational cost.

\section{Experimental Setup}
\label{sec:exp}

\subsection{Datasets and Implementation}

We evaluate VM-ArrayDPS on the SMS-WSJ corpus \cite{drude2019sms}, a spatialized multi-speaker speech corpus designed for multi-channel speech separation and recognition. 
The $2$-speaker test set contains $1,332$ mixtures sampled at $8$ kHz. The mixtures are simulated using a six-microphone circular array with a diameter of $20$ cm. 
The source-to-array distance is uniformly sampled from $1.0$ to $2.0$ m, the reverberation time $T_{60}$ ranges from $0.2$ to $0.5$ s, and weak additive white noise is introduced with an SNR between $20$ and $30$ dB.
Rather than using the complete six-channel array, we select a subset of physical microphones as the observed inputs, following the setting of ArrayDPS and simulating practical scenarios where only a limited number of microphones are available.
Specifically, three physical microphones are used in the $2$-speaker experiments and four physical microphones are used in the $3$-speaker experiments. 
We additionally evaluate VM-ArrayDPS on a $3$-speaker extension generated following the SMS-WSJ simulation procedure.

The pretrained speech diffusion prior and EDM sampling configuration are inherited from the public ArrayDPS implementation \cite{xu2025arraydps}. 
No additional supervised training is performed for VM-ArrayDPS. 
Given the physical microphone observations, IVA is first applied to obtain source-oriented estimates. 
These estimates are then used to construct virtual microphone observations, which are incorporated into the posterior objective together with the physical microphone observations. 
The contribution of the virtual-microphone constraint is controlled by a weighting factor $\alpha$.

For each utterance, we run sampler five times. 
Following ArrayDPS, we consider three sampling-selection strategies. 
``AVG'' averages the five generated outputs, ``Max'' selects the output with the highest SI-SDR, and ``ML'' selects the output with the highest likelihood score. 
We evaluate all three strategies in both the $2$- and $3$-speaker experiments to examine whether the effect of virtual microphones is consistent across different sample-selection procedures.

We report SDR \cite{vincent2006performance}, SI-SDR \cite{le2019sdr}, PESQ \cite{rix2001perceptual}, and eSTOI \cite{taal2011algorithm}. 
PESQ is an objective speech-quality measure standardized in ITU-T P.862, while eSTOI is an extended short-time objective intelligibility measure designed for intelligibility evaluation under challenging distortions \cite{taal2011algorithm}. 
Since the diffusion prior is not explicitly optimized to remove reverberation, the evaluation targets are the corresponding reverberant speaker images. 
Therefore, the reported results measure separation quality with respect to the reverberant targets rather than dereverberation performance.

\section{Evaluation Results}
\label{sec:main_results}

\subsection{Comparison with Existing Methods}

Table \ref{tab:2spk} presents the results on the $3$-channel $2$-speaker SMS-WSJ test set. 
Results marked with $\dagger$ are obtained from previous work or existing reported results and are not re-evaluated in this work.
For a controlled comparison, all VM-ArrayDPS configurations in the table use two virtual microphones with $\alpha=0.5$. 
The effects of the number of virtual microphones and $\alpha$ are investigated separately in Section \ref{sec:ablation}.

VM-ArrayDPS consistently improves over the corresponding ArrayDPS configuration under all three sampling-selection strategies. 
With AVG selection, VM-ArrayDPS improves SDR from $15.8$ to $16.9$ dB and SI-SDR from $15.0$ to $16.1$ dB. PESQ increases from $3.37$ to $3.45$, while eSTOI increases from $0.864$ to $0.881$. 
With Max selection, SDR improves from $17.2$ to $17.9$ dB and SI-SDR from $16.5$ to $17.2$ dB. 
The corresponding PESQ and eSTOI improvements are from $3.52$ to $3.56$ and from $0.888$ to $0.898$, respectively. 
Under ML selection, VM-ArrayDPS improves SDR from $16.9$ to $17.7$ dB and SI-SDR from $16.2$ to $17.0$ dB, while PESQ increases from $3.49$ to $3.53$ and eSTOI from $0.883$ to $0.894$.

The consistent gains across AVG, Max, and ML indicate that the benefit of virtual microphones is not specific to a particular sample-selection strategy. 
Instead, the additional source-oriented observations provide complementary constraints to the DPS process. 
Among the three configurations, the Max configuration achieves $17.9$ dB SDR and $17.2$ dB SI-SDR.

For reference, the conventional spatial clustering and IVA baselines provide lower separation performance than the diffusion-based methods.
IVA with a Gaussian prior reaches $12.2$ dB SI-SDR, while the reported UNSSOR result reaches $14.4$ dB SI-SDR. 
These comparisons provide reference points from both classical spatial separation and unsupervised neural separation, while the direct ArrayDPS comparison isolates the contribution of the proposed virtual-microphone constraints.

\begin{table}[t]
\scriptsize
\sisetup{
    round-mode=places,
    round-precision=2,
    table-number-alignment=center,
    detect-weight=true,
    detect-inline-weight=math
}
\centering
\caption{Results on 3-channel 2-speaker SMS-WSJ test set. Results marked with $\dagger$ are obtained from existing studies. All VM-ArrayDPS results use two virtual microphones with $\alpha=0.5$.}
\label{tab:2spk}
\setlength{\tabcolsep}{5pt}
\renewcommand{\arraystretch}{0.92}
\begin{tabular}{
l
    S[table-format=2.1,round-precision=1]
    S[table-format=2.1,round-precision=1]
    S[table-format=1.2,round-precision=2]
    S[table-format=1.3,round-precision=3]
}
\toprule
Method & {SDR (dB)$\uparrow$} & {SI-SDR (dB)$\uparrow$} & {PESQ$\uparrow$} & {eSTOI$\uparrow$} \\
\midrule
Mixture & 0.1 & 0.0 & 1.87 & 0.603 \\
\midrule
Spatial Clustering$^\dagger$ & 9.5 & 8.5 & 2.52 & 0.759 \\
IVA-Laplace$^\dagger$ & 12.0 & 10.7 & 2.67 & 0.802 \\
IVA-Gaussian$^\dagger$ & 13.4 & 12.2 & 2.82 & 0.834 \\
UNSSOR$^\dagger$ & 15.4 & 14.4 & 3.20 & 0.875 \\
\midrule
ArrayDPS-AVG & 15.8 & 15.0 & 3.37 & 0.864 \\
VM-ArrayDPS-AVG & 16.9 & 16.1 & 3.45 & 0.881 \\
\midrule
ArrayDPS-Max & 17.2 & 16.5 & 3.52 & 0.888 \\
VM-ArrayDPS-Max & \bfseries 17.9 & \bfseries 17.2 & \bfseries 3.56  & \bfseries 0.898 \\
\midrule
ArrayDPS-ML & 16.9 & 16.2 & 3.49 & 0.883 \\
VM-ArrayDPS-ML & 17.7 & 17.0 & 3.53 & 0.894 \\
\bottomrule
\end{tabular}
\end{table}


\begin{table}[t]
\scriptsize
\sisetup{
    round-mode=places,
    round-precision=2,
    table-number-alignment=center,
    detect-weight=true,
    detect-inline-weight=math
}
\centering
\caption{Results on 4-channel 3-speaker SMS-WSJ test set.}
\label{tab:3spk}
\setlength{\tabcolsep}{5pt}
\renewcommand{\arraystretch}{0.92}
\begin{tabular}{l
    S[table-format=2.1,round-precision=1]
    S[table-format=2.1,round-precision=1]
    S[table-format=1.2,round-precision=2]
    S[table-format=1.3,round-precision=3]}
\toprule
Method & {SDR (dB)$\uparrow$} & {SI-SDR (dB)$\uparrow$} & {PESQ$\uparrow$} & {eSTOI$\uparrow$} \\
\midrule
Mixture & -3.0 & -3.2 & 1.57 & 0.458 \\
\midrule
ArrayDPS-AVG & 11.5 & 10.5 & 2.82 & 0.763 \\
VM-ArrayDPS-AVG & \bfseries 13.2 & \bfseries 12.1 & \bfseries 2.95 & \bfseries 0.798 \\
\midrule
ArrayDPS-Max & 13.1 & 12.2 & 3.00 & 0.798 \\
VM-ArrayDPS-Max & \bfseries 14.1 & \bfseries 13.2 & \bfseries 3.07 & \bfseries 0.822 \\
\midrule
ArrayDPS-ML & 12.8 & 11.9 & 3.00 & 0.792 \\
VM-ArrayDPS-ML & \bfseries 13.9 & \bfseries 13.0 & \bfseries 3.05 & \bfseries 0.817 \\
\bottomrule
\end{tabular}
\end{table}

\begin{table}[t]
\scriptsize
\sisetup{
    round-mode=places,
    round-precision=2,
    table-number-alignment=center,
    detect-weight=true,
    detect-inline-weight=math
}
\centering
\caption{Ablation study of number of virtual microphones and virtual-microphone loss weight $\alpha$ on 3-channel 2-speaker SMS-WSJ test set. All results use AVG sample aggregation.}
\label{tab:ablation}
\setlength{\tabcolsep}{4.5pt}
\renewcommand{\arraystretch}{0.92}
\begin{tabular}{
    S[table-format=1,round-precision=0]
    S[table-format=1.1,round-precision=1]
    S[table-format=2.1,round-precision=1]
    S[table-format=2.1,round-precision=1]
    S[table-format=1.2,round-precision=2]
    S[table-format=1.3,round-precision=3]
}
\toprule
{\#virtual microphones} & {$\alpha$} & {SDR (dB)$\uparrow$} & {SI-SDR (dB)$\uparrow$} & {PESQ$\uparrow$} & {eSTOI$\uparrow$} \\
\midrule
0 & {--} & 15.8 & 15.0 & 3.37& 0.864\\
\midrule
2 & 1.0 & 17.0 & 16.1 & 3.44 & 0.883 \\
2 & 0.5 & 16.9 & 16.1 & 3.45 & 0.881 \\
\midrule
6 & 1.0 & 17.0 & 16.1 & 3.43 & 0.882 \\
6 & 0.5 & 17.0 & 16.2 & 3.45 & 0.881 \\
6 & 0.1 & 16.3 & 15.5 & 3.41 & 0.868 \\
\bottomrule
\end{tabular}
\end{table}

We further evaluate VM-ArrayDPS in the $4$-channel $3$-speaker setting, where the separation problem is more under-determined than the $2$-speaker case. 
Table \ref{tab:3spk} compares ArrayDPS and VM-ArrayDPS under the same three sampling-selection strategies.
With AVG selection, VM-ArrayDPS improves SDR from $11.5$ to $13.2$ dB and SI-SDR from $10.5$ to $12.1$ dB, corresponding to gains of $1.7$ and $1.6$ dB, respectively. 
PESQ increases from $2.82$ to $2.95$, while eSTOI increases from $0.763$ to $0.798$. 
With Max selection, SDR increases from $13.1$ to $14.1$ dB and SI-SDR from $12.2$ to $13.2$ dB. 
PESQ and eSTOI also improve from $3.00$ to $3.07$ and from $0.798$ to $0.822$, respectively.
The same trend holds for ML selection. 
VM-ArrayDPS improves SDR from $12.8$ to $13.9$ dB and SI-SDR from $11.9$ to $13.0$ dB, while PESQ increases from $3.00$ to $3.05$ and eSTOI from $0.792$ to $0.817$. 
Thus, the proposed virtual-microphone constraint provides consistent improvements across all three sampling-selection strategies even when the number of sources increases.

\subsection{Ablation Results}
\label{sec:ablation}

We investigate the effects of the number of virtual microphones and the virtual-microphone loss weight $\alpha$. We use the $3$-channel $2$-speaker setting with AVG sampling to isolate these factors. The results are summarized in Table \ref{tab:ablation}.

We first examine the number of virtual microphones. Adding $2$ virtual microphones substantially improves the ArrayDPS baseline. With $\alpha=1.0$, SDR increases from $15.8$ to $17.0$ dB and SI-SDR from $15.0$ to $16.1$ dB. Increasing the number of virtual microphones from $2$ to $6$ provides no further improvement: both configurations obtain $17.0$ dB SDR and $16.1$ dB SI-SDR. A similar trend is observed with $\alpha=0.5$, where SI-SDR changes only from $16.1$ to $16.2$ dB. This suggests that a small number of virtual microphones is sufficient, as additional channels derived from the same physical observations may provide increasingly redundant information.

We next examine the effect of $\alpha$. With two virtual microphones, reducing $\alpha$ from $1.0$ to $0.5$ keeps SI-SDR at $16.1$ dB and slightly improves PESQ from $3.44$ to $3.45$. With six virtual microphones, $\alpha=0.5$ gives a slightly higher SI-SDR of $16.2$ dB than $16.1$ dB at $\alpha=1.0$, with a small improvement in PESQ. Further reducing $\alpha$ to $0.1$ causes clear degradation, with SI-SDR dropping to $15.5$ dB and eSTOI to $0.868$. These results suggest that $\alpha=0.5$ provides sufficient weight for the virtual-microphone constraint, while further reducing $\alpha$ weakens its contribution and degrades performance.

Based on these results, we use $2$ virtual microphones with $\alpha=0.5$ as the default configuration in the main experiments. The three sampling-selection strategies are then evaluated independently, as reported in Table \ref{tab:2spk} and \ref{tab:3spk}.

\section{Conclusions}
\label{sec:conclusion}

We have proposed VM-ArrayDPS, a virtual-microphone augmentation method for unsupervised BSS with diffusion posterior sampling. By reusing higher-SNR IVA outputs as virtual microphone observations, the proposed method introduces additional multi-channel consistency constraints and provides cleaner observations for FCP-based acoustic estimation. On both 2- and 3-speaker SMS-WSJ, VM-ArrayDPS consistently improves over ArrayDPS across multiple sampling-selection strategies.
Ablation results further shows that a small number of virtual microphones is sufficient and that assigning them a moderate consistency weight is preferable to treating them as perfectly reliable measurements. These results indicate that virtual observations provide a simple and effective way to strengthen the measurement side of diffusion-based BSS without additional supervised training.

{\footnotesize
\bibliographystyle{IEEEtran}
\bibliography{references}
}

\end{document}